# Spectroscopic Determination of Site-Selective Ligand Binding on Single Anisotropic Nanocrystals

Dong Le,[†,‡,⊥] Wade Shipley,[¶,‡,⊥] Alexandria Do,[¶,‡,⊥] Liya Bi,[§,‡,⊥] Yufei Wang,[¶,‡] Krista P. Balto,[§] Rourav Basak,[†] Hans A. Bechtel,[∥] Stephanie N. Gilbert Corder,[∥] Ilya Mazalov,[∥] Tesa Manto,[†] Reno Sammons,[†] Yutong She,[¶] Fiona Liang,[¶] Ganesh Raghavendran,[¶] Joshua S. Figueroa,[§] Shaowei Li,[§,‡] Tod A. Pascal,[¶,‡] Andrea R. Tao,[¶,‡] and Alex Frano*[,†,‡]

[†]*Department of Physics, University of California, San Diego, La Jolla, CA, USA.*
[‡]*Materials Science and Engineering, University of California, San Diego, La Jolla, CA, USA*
[¶]*Aiiso and Yufeng Li Family Department of Chemical and NanoEngineering, University of California, San Diego, La Jolla, CA, USA.*
[§]*Department of Chemistry and Biochemistry, University of California, San Diego, La Jolla, CA, USA.*
[∥]*Advanced Light Source, Lawrence Berkeley National Laboratory, Berkeley, CA, USA.*

**ABSTRACT:** Organic surface ligands are integral components of nanocrystals and nanoparticles that have a strong influence on their physicochemical properties, their interaction with the environment, and their ability to self-assemble and order into higher-order structures. These hybrid nanomaterials are tunable with applications in catalysis, directed self-assembly, next-generation optoelectronics, and chemical and quantum sensing. Critically, future advances depend on our ability to rationally engineer their surface chemistry. However, fundamental knowledge of ligand-nanoparticle behavior is limited by uncertainty in where and how these ligands bind to surfaces. For nanoparticles, in particular, few characterization techniques offer both the high spatial resolution and the precise chemical mapping needed to identify specific ligand binding sites. In this study, we utilized synchrotron infrared nanospectroscopy (SINS), atomic force microscopy (AFM), and scanning tunneling microscopy (STM) together with first-principles computer simulations to validate the site-selective adsorption of organic ligands on a shaped nanocrystal surface. Specifically, we demonstrate that the sterically encumbered isocyanide ligands ($CNAr^{Mes2}$) preferentially bind to the high curvature features of Ag nanocubes (NCs), where low-coordinate Ag atoms are present. In contrast, isocyanide ligands that do not exhibit these steric properties show no surface selectivity. SINS serves as an effective tool to validate these surface binding interactions at the near-single molecule level. These results indicate that steric effects can be successfully harnessed to design bespoke organic ligands for fine-tuning nanocrystal surface chemistry and the properties of the nanocrystal ligand shell.

## INTRODUCTION

In nature, selective chemical interactions, such as those between enzymes and their substrates, occur with remarkable precision, enabling complex biological processes to proceed efficiently and accurately.[1,2] For years, researchers have sought to emulate lock-and-key molecular recognition by designing ligands,[3,4] engineering surface chemistry,[5] or employing molecular templates[6] to precisely control interactions at the nanoscale. The concept of selective binding extends naturally to the chemical topography of colloidal nanoparticles, where emulating molecular recognition has the potential to govern the crucial ligand-nanoparticle interactions that dictate nanoparticle stability and function.[7,8] To this end, a promising approach is the rational design of organic ligands tailored for inorganic NPs, in which systematic modification of the headgroup and tail has been shown to tune surface chemistry and improve performance. For example, Morad *et al.* demonstrated that phospholipids with modified head and tail groups can functionalize lead halide perovskite nanocrystals, yielding improved colloidal stability, solvent compatibility, and optoelectronic properties.[4] Similar approaches have been applied to phosphonic acids for metal oxide nanoparticles,[9] thiols for noble metal nanocrystals,[3] and zwitterionic ligands for improved stability in polar media.[10] Despite these advances, relatively few studies have examined how head group topography influences site-specific surface recognition. Notably, many widely used ligands, such as trioctylphosphine oxide (TOPO), are known to adopt multiple binding configurations on quantum dot nanoparticle surfaces,[11] underscoring the complexity of ligand–surface interactions. Building on this, we propose that using designer organic ligands that capable of selectively binding to *specific sites* on NP surface can be highly beneficial.

Among the many factors that influencing the binding behavior of ligands on NP surfaces, steric effects play an important role by dictating the binding geometry of the ligand molecules and the way in which they coordinate on the metal surface atoms.[12–14] A good example of the steric effect is the use of bulky ligands to modify the surface of metal NPs.[15,16] Ligands with sterically encumbering organic substituents,[17–20] such as *m*-terphenyl isocyanide ligands,[21–23] are promising due to their ability to interact with various transition metal centers through the isocyanide group (Fig. 1a-b). However, the encumbering aromatic backbone imposes substantial steric constraints (Fig. 1c), which enforce and help stabilize low coordination numbers for the transition metal center through steric interference.[20,22,24] These steric effects make such ligands well-suited for tuning metal-ligand interactions and improving the overall stability of metal complexes.[25–30] Furthermore, these ligands exhibit selective binding affinity with nanocrystals, especially in high-curvature regions.[31–33] Using a series of *m*-terphenyl isocyanide ligands, Wang *et al.* showed that ligand-surface steric pressures are maximized on planar metal surfaces, but significantly reduced on convex surfaces, such as step edges, where the *m*-terphenyl group experiences less steric hindrance.[31] Although previous studies have focused primarily on spherical NPs, the influence of steric hindrance on anisotropic shapes, particularly faceted polyhedral shapes common in the literature, remains less explored. In this study, we examine these effects using Ag nanocubes (AgNCs) (Fig. 1d) to provide new insight into how ligand-surface steric interactions modulate ligand-NP binding, and more specifically govern selective surface recognition.

Critical to this study are characterization techniques that provide molecular-level spatial resolution and detailed chemical information of the NP surface, simultaneously; however, most conventional techniques typically trade one for the other. Techniques such as high resolution transmission electron microscopy (HR-TEM) can provide the atomic-level surface structure of NPs and in rare instances identify bound ligands.[34] Further, atomic force microscopy (AFM) can provide topography of ligated NPs,[35] and nuclear magnetic resonance (NMR) spectroscopy can provide information on NP surface chemistry,[36] but integration of these capabilities to achieve high-resolution selection of ligands on surfaces remains rare. To overcome this, we employ synchrotron infrared nanospectroscopy (SINS) to directly probe the binding locations and chemical signatures of IR-active isocyanide ligands on a single nanocube (NC). SINS is a near-field technique that incorporates a commercial rapid-scan Fourier transform infrared (FTIR) interferometer and an AFM into an asymmetric Michelson interferometer set-up.[37–40] By focusing synchrotron infrared (IR) light on a sharp modified AFM tip, which acts as an antenna to significantly enhance the IR field near its apex, SINS can achieve a spatial resolution of around 20 nm, well below the diffraction limit and independent of the wavelength used.[37–40] This makes SINS particularly suitable for studying ligand-NP interfaces, as it provides spatial and chemical information at the same time. Previous studies have used SINS to detect and map molecular vibrations on NPs surfaces, demonstrating its ability to probe surface chemistry at the nanoscale.[41–44]

Here, we use SINS as a linchpin tool for the experimental characterization of ligand binding behavior on the surface of shaped AgNCs together with scanning tunneling microscopy (STM) and theoretical predictions. Quantum mechanical (QM) electronic-structure calculations and molecular dynamics (MD) simulations suggest that the encumbered *m*-terphenyl isocyanide ligand, CNAr$^{Mes2}$ (where Ar$^{Mes2}$ = 2,6-(2,4,6-Me$_3$C$_6$H$_2$)$_2$C$_6$H$_3$); Fig. 1a) preferentially binds to convex regions or sites with low-coordinate Ag atoms, whereas the less sterically hindered isocyanide, CNXylyl (where Xylyl = (CH$_3$)$_2$C$_6$H$_3$; Fig. 1b) show no strong site preference. Point SINS measurements taken at both the corners and center of individual NCs provide direct, unambiguous support for these predictions: CNAr$^{Mes2}$ ligands are predominantly observed near the corners, while a non-binding vibrational mode, attributed to excess unbound ligands, is present at both locations. In contrast, CNXylyl-AgNCs exhibit ligand binding signals across all surface positions, consistent with the predicted lack of site selectivity. The vibrational binding mode of a single CNAr$^{Mes2}$ molecule on the Ag surface is confirmed by scanning tunneling microscopy (STM) and STM-inelastic electron tunneling microscopy (STM-IETS).[45–47] Taken together, these results demonstrate that steric hindrance governs ligand binding preferences on anisotropic NP surfaces and highlight the utility of the SINS technique in directly probing site-specific interactions with chemical specificity at the single-particle level. Such insights are important for guiding the rational design of organic ligands to fine-tune the surface chemistry and properties of the nanocrystal ligand shell.

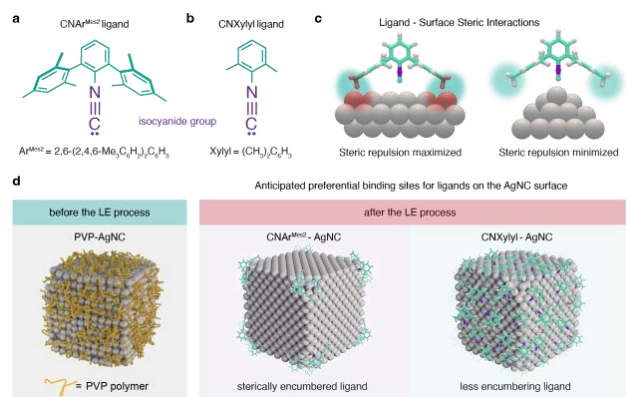

**Figure 1.** Schematic illustrating how steric effect can be utilized to design ligand with target specific. Two types of isocyanide ligands used in this study are compared based on their structural differences: (a) the sterically encumbered *m*-terphenyl isocyanide CNAr$^{Mes2}$, with three aromatic rings forming a cumbersome backbone, and (b) the less sterically protective CNXylyl ligand, which has a single aromatic ring. (c) Schematic representation of steric effect showing how steric pressure is maximized on planar surface (left) due to limited spatial accommodation compared to step-edge site (right). (d) Illustration of anticipated preferential ligand-binding sites on the nanocube (NC) surface, before and after the ligand exchange (LE) process.

## RESULTS AND DISCUSSION

**Predicted isocyanide ligand binding preferences using QM and MD simulations.** To understand how molecular structure influences the binding of ligands on Ag surfaces, we performed QM calculations using density functional theory (DFT) at the PBE/GGA level with dispersion corrections (see Computational Methods Section). Two representative Ag surfaces were modeled: a perfectly flat Ag slab, analogous to the center planar regions of the cube, and an Ag step-edge site, which resembles the local geometry at the NC corners. The binding energy was defined as the minimum interaction energy corresponding to the closest approach of the ligand to the Ag surface. In vacuum, CNAr$^{Mes2}$ binds more strongly to the step-edge sites, with a minimum interaction energy of $E = -0.87$ eV at an Ag–C bond distance of 2.1 Å (Fig. 2a), consistent with previously reported experimental values for the silver-carbon bond distances.[48] However, on the flat Ag surface, the binding energy is significantly weaker ($E = -0.19$ eV), with a longer Ag–C bond distance of 3.1 Å (Fig. 2a), indicating reduced orbital overlap and thus weaker metal–ligand binding, or perhaps no binding at all. We reasoned that these differences are caused by the bulky mesityl wings of the CNAr$^{Mes2}$ molecule, which limit the ligand's ability to approach the surface closely. We also carried out a comparative study with a less hindered isocyanide ligand, CNXylyl, to further demonstrate the relationship between steric hindrance and ligand binding patterns (Fig. 1b). Unlike CNAr$^{Mes2}$, CNXylyl maintains similar binding distances (~2.1 Å) at both the step edge and the flat surface (Fig. 2b). Although it still binds more strongly to the edge site (−0.89 eV vs. −0.59 eV), the relative difference in the binding energy between the edge and center sites is modest (~50%) compared to that of CNAr$^{Mes2}$ (~350%).

The reduced sensitivity to surface geometry observed for CNXylyl can be attributed to its smaller steric profile, which allows easier access to the surface and minimizes the energetic cost associated with binding to flat, planar regions.

Next, we performed atomistic MD simulations of an AgNC suspended in a solution of chloroform and CNAr$^{Mes2}$ molecules. The ligand–Ag interaction potentials were parameterized using binding energy profiles obtained from our DFT calculations (see Computational Methods section). The particle was modeled as a cube with edge lengths of 32.68 Å, bounded by Ag(100) and Ag(111) facets (Fig. S17). As we previously demonstrated on Au nanosphere,[31] the presence of solvent molecules modulates ligand adsorption to the Ag surface, which we explored through extensive MD simulation trajectories. The final snapshot of the system (with solvent molecules omitted for clarity) and the statistical analysis of the distribution of the ligands on a single cube reveal that 85% of CNAr$^{Mes2}$ ligands preferentially bind to the corners or edges of the AgNC, where low-coordinate Ag atoms are abundant (Fig. 2c). These corner and edge sites provide more spatial accommodation for the bulky mesityl groups of CNAr$^{Mes2}$, resulting in preferential binding in regions of higher curvature. Throughout the simulation trajectory, no desorption events were observed, consistent with the strong ligand– Ag interactions predicted by DFT. Analogous MD simulations using CNXylyl ligands under the same conditions demonstrated that in contrast to CNAr$^{Mes2}$, the CNXylyl ligands exhibited a more uniform binding pattern: more than 90% of the bound ligands were found in the flat planar facets and corners of the cube (Fig. 2d). This uniform surface distribution can be attributed to the less encumbered steric profile of CNXylyl, which allows access to curved and flat surface sites with minimal energetic penalty. The binding trends observed in the MD simulations in both cases aligned well with our DFT results and provided a consistent theoretical framework to guide our experimental investigations.

**Experimental validation of ligand binding patterns on individual nanocubes probed by SINS.** Building on our theoretical predictions, we used SINS to experimentally investigate the site-selective binding behavior of isocyanide ligands on a single-shaped AgNC. Two reasons make SINS particularly advantageous for this study: first, the isocyanide functional group is IR-active, allowing us to distinguish between free and Ag-bound species based on shifts in the C≡N stretching frequency; second, the method is compatible with our AFM-based sample preparation, allowing measurements at the single-particle level. As shown in Fig. 3a, the synchrotron IR near-field achieves a spatial resolution of approximately 20 nm,[37,39,40] sufficient to resolve binding differences between the distinct regions of our individual AgNCs, which are roughly 80 nm in size.

We prepared the AgNC samples for SINS measurements through a ligand exchange (LE) process, in which the native polyvinylpyrrolidone (PVP) polymer was replaced with either CNAr$^{Mes2}$ or CNXylyl. The ligand-grafted AgNCs were then deposited onto an Au substrate using a dip coating method to form a uniform particle film. The resulting spectra are normalized by the Au response using custom software developed at Beamline 5.4 of the Advanced Light Source at Lawrence Berkeley National Laboratory (LBNL). Point SINS measurements were then performed at selected locations on individual CNAr$^{Mes2}$–ligated AgNCs, with the tip positioned at regions near the corner (red) and the center (dark gray) of the AgNC, as indicated by solid circles in the AFM image (Fig. 3b). The corresponding IR vibrational spectra collected at these sites are shown in Fig. 3c. To minimize instrumental effects and improve signal-to-noise ratios, we collected multiple interferograms at each location and averaged them using established procedures.[37,41,42] The vibrational modes in the C≡N

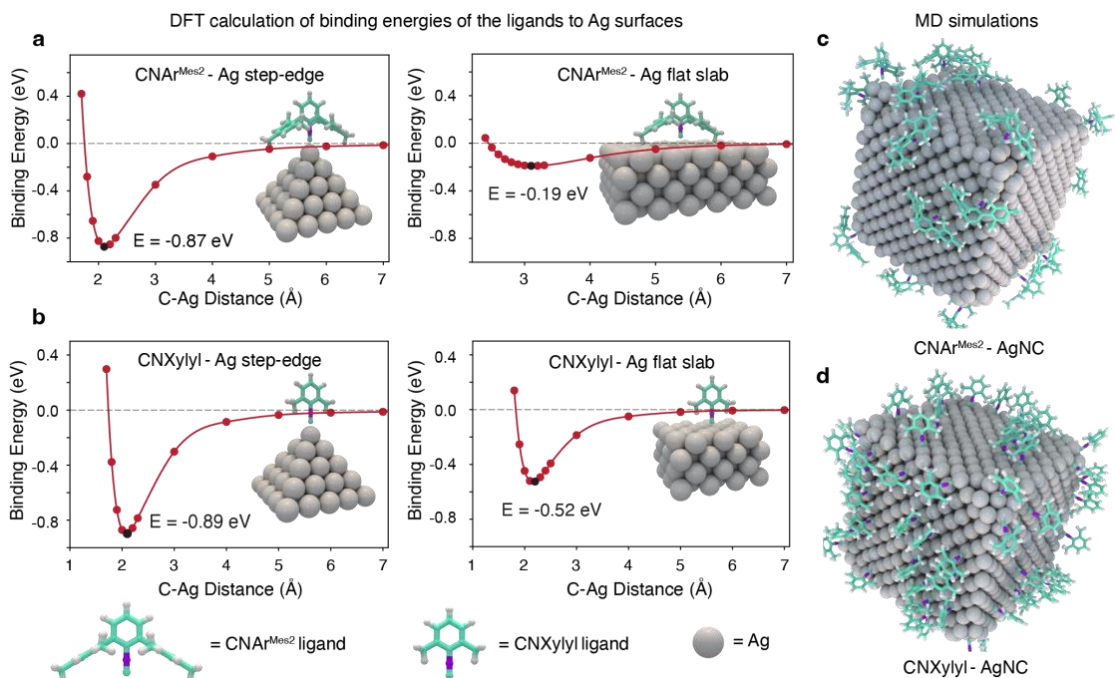

**Figure 2.** QM and MD simulations predicted increasing the steric hindrance of isocyanide ligands directs them to specific binding sites. DFT interaction energy plots for CNAr$^{Mes2}$ (a) and CNXylyl (b) binding to an Ag step-edge site (left) and a flat planar Ag surface (right). The data (solid red circles) are connected by lines to guide the eye, with solid black circles indicating the most favorable energy. (c-d) Final snapshots from 5 ns MD simulations of CNAr$^{Mes2}$ and CNXylyl ligands on an Ag nanocube, respectively, with chloroform solvent molecules omitted for clarity

stretching region were assigned based on our ATR-FTIR measurements (Fig. S1 of the Supporting Information) and previous studies.[21–23] A peak at 2114 cm$^{-1}$ corresponds to the stretching mode of free or physisorbed isocyanide ligands, while a higher-frequency peak at 2150 cm$^{-1}$ is attributed to chemisorbed CNAr$^{Mes2}$ bound to Ag atoms on the AgNC surface. The schematic representations of these two binding modes are shown in Fig. 3d.

To analyze the spectra more quantitatively, we used Gaussian peak fitting to deconvolute the C≡N stretch region for both corner and center spectra and detailed results are presented in Supplementary Figure S15. The assignment of peaks was based on the dominant intensities of the physisorbed $v_s$(C≡N) and chemisorbed $v_s$(Ag–C≡N) modes. Given the relatively low spectral resolution and signal-to-noise ratio of the SINS spectra, it is not straightforward to assign those peaks that have lower intensities compared to $v_s$(C≡N) and $v_s$(Ag–C≡N). We suspect some of these splitting in the C≡N stretch region may arise from intermolecular interactions or arise from the orientation variation[37,40,43,49] of the ligand on the AgNC surface with a large curvature, which enhances near-field coupling. The probe is also sensitive to the out-of-sample-plane dipole component (perpendicular to the sample surface) and therefore preferentially detects these orientations. However, a complete interpretation will require more refined theoretical models that explicitly incorporate the tip–sample interaction, which is quite challenging. Regardless, our primary goal was to identify the location of ligand binding on the cube surface after the ligand exchange (LE) process, which the data clearly support. The peak at $v_s$(C≡N) = 2114 cm$^{-1}$ appears in both corner and center spectra, consistent with a residual population of physisorbed CNAr$^{Mes2}$ remaining on the surface. In contrast, we can see a distinct peak at $v_s$(Ag–C≡N) = 2150 cm$^{-1}$ in the corner spectrum, indicating the chemisorption of CNAr$^{Mes2}$ at this location. These observations agree with our DFT and MD results, which predict preferential binding of bulky CNAr$^{Mes2}$ ligands at high curvature and low coordinate Ag atoms. It should be noted that the blue shift in frequency from physisorption to chemisorption is attributed to the donation of $\sigma$-electrons from the isocyanide to Ag in a terminal $\eta^1$ bond mode, rather than the donation of electrons from Ag to isocyanide via π-backbonding.[50–52] In complexes featuring metals that are weakly π-basic, isocyanide σ-donation to the metal relieves the antibonding character in the C≡N triple bond, leading to blue shifted $v_s$(M–C≡N) vibrations relative to free isocyanide.[50–52] Moreover, this blue-shift behavior is consistent with observations of isocyanides adsorbed on other forms of Ag, including films and powders, and is consistent with other studies of isocyanide coordination chemistry on Au.[51–55]

Next, we examined how reduced steric hindrance influences the ligand binding behavior using CNXylyl, a ligand with a smaller head group containing only a single aromatic ring (Fig. 1b). Previous studies on Au nanospheres have shown that CNXylyl binds indiscriminately to all surface sites and does not select only for regions of high nanocurvature.[31] Following the same approach as with CNAr$^{Mes2}$, point SINS measurements were performed on individual CNXylyl–AgNCs at two positions: near the corner (red) and the center (dark gray), as marked in the AFM image (Fig. 3e). Our ATR-FTIR measurements (Supplementary Fig.S2) indicate that the C≡N stretching frequency of CNXylyl bound to Ag appears at approximately 2174 cm$^{-1}$. In the SINS spectra, a peak at 2167 cm$^{-1}$ was observed at both corner and center locations (Fig. 3f), which we assign to the Ag–isocyanide stretching mode, $v_s$(Ag–C≡N). This vibrational mode appears at a higher frequency than the chemisorbed peak for CNAr$^{Mes2}$–AgNCs (2149 cm$^{-1}$), a trend that is consistent with earlier studies showing that CNXylyl generally exhibits higher C≡N stretching frequencies than CNAr$^{Mes2}$, despite their similar electronic properties.[53] Specifically in solution, CNXylyl absorbs at 2122 cm$^{-1}$, while CNAr$^{Mes2}$ is observed at 2112 cm$^{-1}$.[53] The presence of 2167 cm$^{-1}$ peaks at both the corner and center positions indicates that the CNXylyl ligands bind across the NC surface without a clear preference for specific sites. This experimental result also aligns with our previous DFT and MD simulations, which showed only a small energy difference for CNXylyl binding at step-edge versus planar surfaces and a more uniform ligand distribution across the cube surface. Together, these findings suggest that in the absence of significant steric constraints, CNXylyl ligands readily access both low- and high-curvature sites, leading to a non-selective binding pattern.

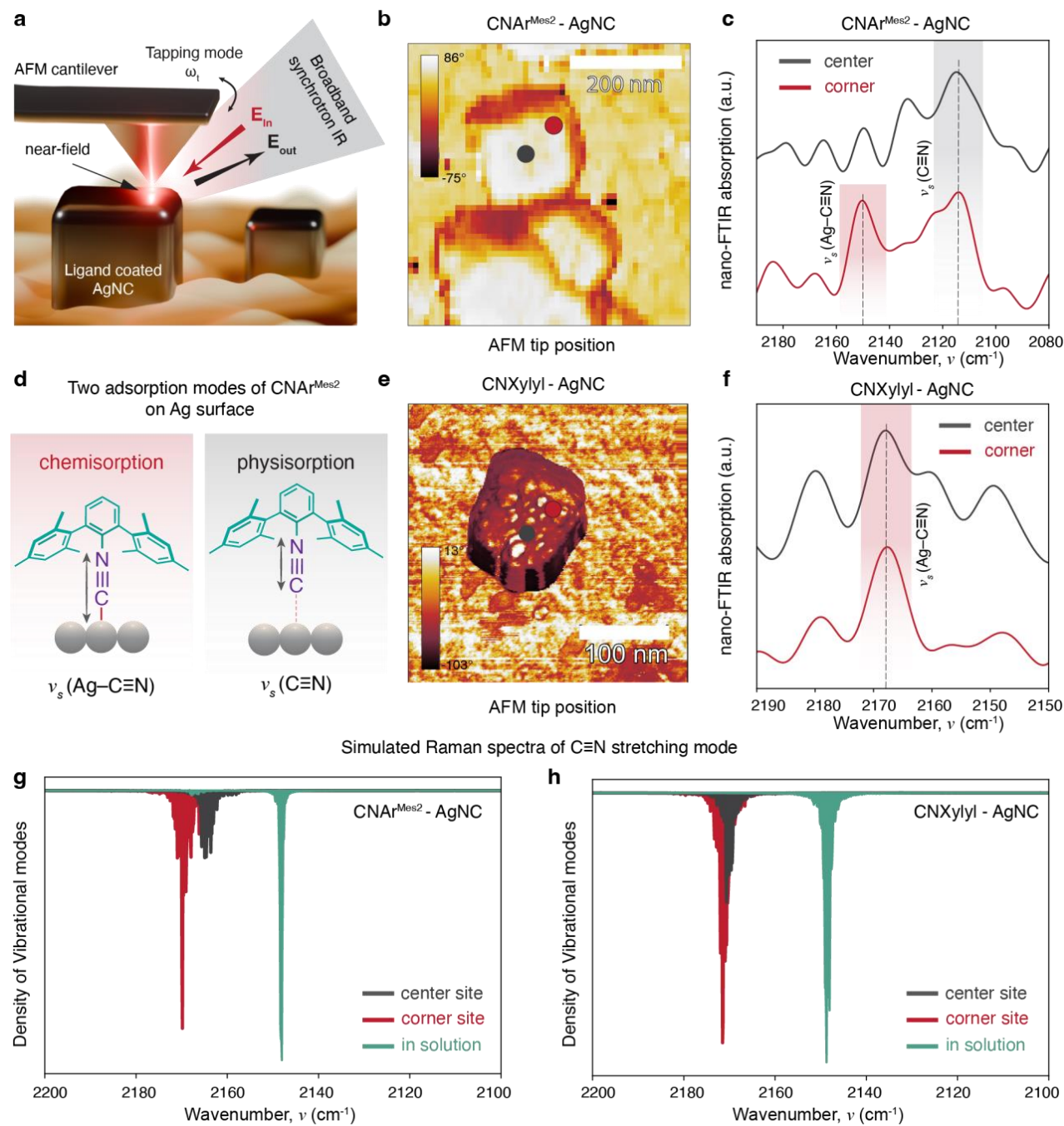

**Figure 3.** Near-field nano-FTIR vibrational spectroscopy validating ligand binding preferences on a single anisotropic AgNC. (a) Close-up schematic of the SINS experiment. A PtSi-coated AFM tip, oscillating in tapping mode at a $\omega_t$ frequency, functions as an optical antenna, focusing the broadband synchrotron infrared (IR) light. The interaction of the IR light with the metallic-coated tip creates a strong near-field focus at the apex of the tip. This nanoscale focus enables localized probing of vibrational modes in nanostructures and molecular systems on targeted surfaces. (b) AFM image of CNAr$^{Mes2}$–AgNC showing tip measurement locations, corner (red) and center (black), with corresponding nano-FTIR vibrational spectra in (c). Scale bar, 200 nm. (d) Illustration depicting two isocyanide adsorption modes on Ag that can be detected by IR spectroscopy: chemisorption (left) and physisorption (right). (e) AFM image of CNXylyl–AgNC with corner and center measurement locations, and (f) the corresponding nano-FTIR spectra. Scale bar, 100 nm. (g-h) Simulated Raman spectra of C≡N stretching modes for CNAr$^{Mes2}$–AgNC and CNXylyl–AgNC, respectively, comparing corner, center sites, and unbound ligands.

To investigate how molecular binding geometry influences the C≡N stretching modes, we performed vibrational analysis of the Raman and IR active modes by modifying an approach for generating the vibrational density of states function using autocorrelation functions directly from MD trajectories.[56] For CNAr$^{Mes2}$, the simulated spectra revealed distinct peaks at 2170 and 2165 cm$^{-1}$ for ligands bound at the corner and planar sites, respectively (Fig. 3g). A separate peak at 2148 cm$^{-1}$ corresponds to the unbound ligand and is consistent with previously reported simulation results.[31] These trends mirror our experimental SINS measurements and DFT predictions, in which higher curvature regions exhibit stronger binding interactions with CNAr$^{Mes2}$. In terms of vibrational mode population, the simulated spectra also show a higher density of C≡N modes at the corners compared to the center, supporting the equilibrium MD simulation results that CNAr$^{Mes2}$ preferentially adsorbs at corner and edge sites. We applied the same approach to simulate the C≡N stretching modes of the CNXylyl ligands (Fig. 3h). Here, and in contrast to CNAr$^{Mes2}$, the CNXylyl spectra exhibited only a slight shift between the corner and center sites, with peaks at 2172 and 2171 cm$^{-1}$, respectively. This small difference indicates that the binding geometry has minimal impact on the C≡N stretching frequency for CNXylyl. Again, these results are in line with our DFT and experimental findings, which show no strong preference for binding at highcurvature regions. Furthermore, the simulated spectra reveal a greater density of vibrational modes at the center sites of CNXylyl compared to CNAr$^{Mes2}$, suggesting that a larger proportion of CNXylyl ligands occupy the planar facets of the nanocube. This is consistent with both MD simulation results and the experimentally observed lack of site selectivity for this less-sterically hindered ligand.

We also performed high-resolution AFM characterization on the samples before and after the LE process to emphasize the importance of a spatially resolved chemical mapping technique like SINS (Supplementary Fig.S14). For PVP-AgNCs, the surface appears uniformly coated, with occasional polymer clustering likely due to drying artifacts. After the LE process with CNAr$^{Mes2}$, the surface morphology is rougher, and topographic profiles reveal features on the order of 1 nm— which is the estimated size of a single CNAr$^{Mes2}$ molecule, as opposed to the flatter profile of the initial PVP-grafted AgNCs, excluding the larger PVP clusters. In the CNXylyl-AgNCs sample, AFM images show a more uniform distribution of small clusters, likely a mix of CNXylyl and residual PVP. In both cases, the data suggested that the ligands have replaced some of the native PVP and adsorbed to the AgNC surface. However, while AFM can effectively detect the change in the surface morphology of samples before and after the LE process, it cannot differentiate between chemisorbed and physisorbed states. This result highlights the value of near-field vibrational spectroscopy in probing the interaction between ligands and NP surfaces, providing important information that AFM alone cannot provide.

**Mechanistic insights into the binding of CNAr$^{Mes2}$ to AgNC corners.** To further understand the binding mechanism of CNAr$^{Mes2}$ at AgNC corner sites, we employed scanning tunneling microscopy (STM) to investigate its adsorption behavior on Ag(111) surface, since the (111) facet is commonly used to represent the corner facet surfaces of AgNCs (Fig. 4a-b).[33,57] A schematic of the STM setup is shown in Fig. 4a, where a metal probe tip scans a conducting Ag(111) substrate and measures the tunneling current, $I_t$, in between. The zoomed-in region illustrates the tip positioned a few Angstroms above the Ag surface, allowing for a clear visualization of individual molecular adsorbates. The CNAr$^{Mes2}$-bound Ag(111) sample was prepared following a previously reported procedure for the deposition of CNAr$^{Mes2}$ on coinage metal surfaces.[32] STM imaging shows that the majority of CNAr$^{Mes2}$ molecules preferentially populate the convex surface sites, particularly along the step edges of Ag(111), as highlighted by black arrows in the topographic image shown in Fig. 4c. This observation confirms the preference of CNAr$^{Mes2}$ to bond to curved surface sites and supports the SINS results, where chemisorbed CNAr$^{Mes2}$ ligands were primarily detected near the most corrugated surface regions, (i.e., AgNC corners).

In a previous study detailing the binding of CNAr$^{Mes2}$ to Ag(111) surfaces, we showed that individual isocyanide ligands prefer to bind to Ag adatom present near step edges as a mechanism to release steric strain.[33] This binding mechanism is facilitated by the intrinsic mobility of Ag atoms on the surface at room temperature, which can readily rearrange to accommodate encumbering ligands.[33] A similar adatom-mediated binding mechanism very likely contributes to the adsorption of CNAr$^{Mes2}$ on the NC surfaces studied here, especially at the corners, where many Ag atoms may not be arranged in a perfectly ordered lattice (Fig. 4b). These structural irregularities at the corners could promote the stabilization of CNAr$^{Mes2}$/adatom assembly, facilitating ligand chemisorption in a manner similar to that observed at Ag(111) step edges (Fig. 4c). Moreover, to provide further characterization of the binding environment of individual surface-bound CNAr$^{Mes2}$ molecules near the step edges, we employed STM-inelastic electron tunneling spectroscopy (STM-IETS) . Fig. 4d shows a SINS, which can enable simultaneous spatial and chemical mapping at the single-particle level. SINS revealed distinct representative IETS spectrum acquired over a single CNAr$^{Mes2}$ at an Ag(111) step edge (Fig. 4d inset). IETS reveals a distinct vibrational characteristic at around 266 meV (around 2145 cm$^{-1}$), which we assign to the Ag-bound isocyanide stretching vibration. Importantly, it has been demonstrated by STM-IETS that $v_s$(M–C≡N) bands are very sensitive to the specific electronic environment present for metal-bound isocyanides, and can shift dramatically as a function of surface site, coordination mode and coordination geometry.[58] Accordingly, the fact that $v_s$(M–C≡N) bands are closely matched in both the SINS and the STM-IETS spectra, strongly indicates that the Ag(111) surface interactions serves a reliable representation for CNAr$^{Mes2}$ binding to the (111)-faceted corner of the AgNC.

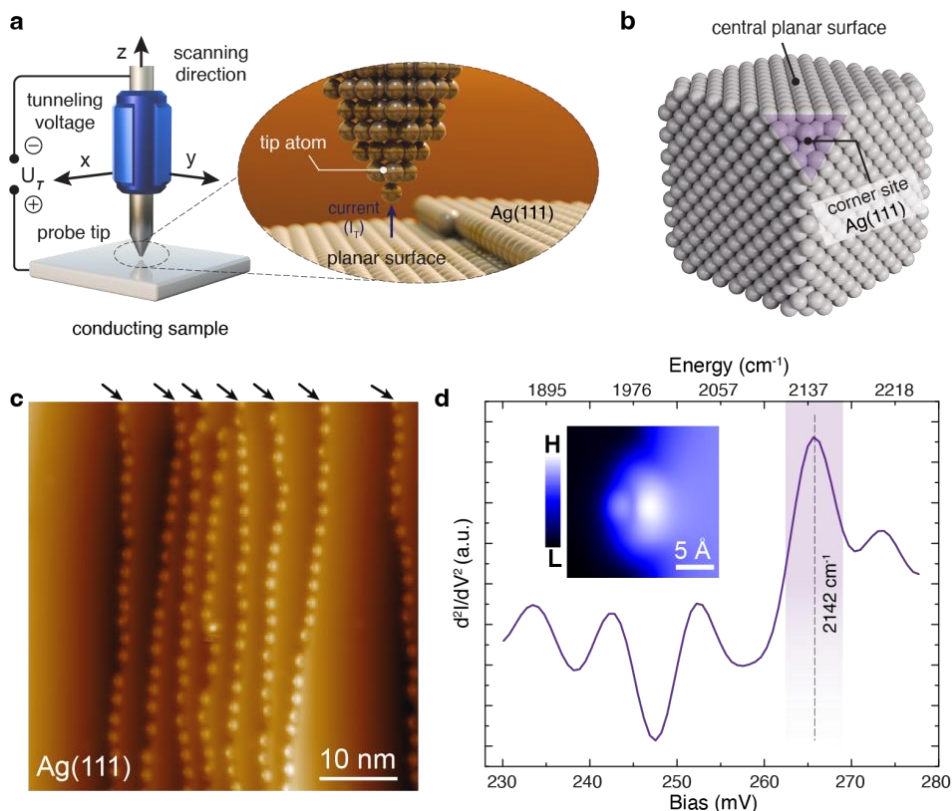

**Figure 4.** Selectively populated high-curvature surface sites with CNAr$^{Mes2}$ and the vibrational binding mode of a single molecule on Ag surface by STM. (a) Schematic of a STM measurement with a metal probe tip scanning a conducting sample. The zoomed-in figure shows a close-up of the tip positioned above the Ag(111) surfaces. (b) Cartoon of an Ag nanocube highlighting the similarity between the Ag(111) step-edge and the NC corner, which is commonly believed to be terminated with (111) facet. (c) Large-scale topographic image of CNAr$^{Mes2}$ molecules on Ag(111). CNAr$^{Mes2}$ predominantly adsorbs close to the step edges (black arrows). (d) IETS spectrum of a CNAr$^{Mes2}$ at the Ag(111) step edge. The peak corresponding to the Ag-bound isocyanide stretching vibration is labeled by the dashed line. Imaging and spectroscopy parameters were -1 V, 50 pA (c), -100 mV, 100 pA (d inset) and -300 mV, 1 nA (d).

## CONCLUSION

In this work, we demonstrated that the combination of computational modeling and multimodal, advanced spectroscopic techniques can elucidate the factors that govern ligand adsorption on shaped metal NP surfaces. Using DFT and MD simulations, we first predicted that steric effects would drive site-selective binding of the encumbering CNAr$^{Mes2}$ ligand to high-curvature regions of AgNCs. In contrast, the unencumbered CNXylyl ligand would exhibit a more uniform surface coverage. These predictions were validated through chemisorption of CNAr$^{Mes2}$ at the AgNC corners, supported by AFM and STM measurements that provided complementary morphological and vibrational evidence for site-specific ligand adsorption. The observed preference for corner binding sites on AgNC is likely driven by an adatom-mediated binding mechanism that relieves steric crowding from the ligand's aryl substituents. Together, these results establish a clear link between ligand structure and surface site preference, offering useful design rules for tuning nanoscale surface functionalization. In the future, our work will focus on engineering the tail groups of CNAr$^{Mes2}$ to dictate the placement and direct the self-assembly of site-selective ligands on AgNCs, facilitating precise chemical functionalization. Although isocyanides exhibit strong interactions with various transition metal surfaces, a deeper understanding of their adsorption geometries and binding determinants could unlock new strategies for designing molecular interfaces and catalytic systems.

## METHODS

**Chemicals and Materials.** Silver nitrate (≥ 99%), 1,5-pentanediol (PD, ≥ 97%), copper (II) chloride (≥ 98%), poly(vinylpyrrolidone) (PVP, average $M_W$ = 55,000 g/mol), Xylyl isocyanide (CNXyl; Xylyl = 2,6-Me$_2$C$_6$H$_3$; 98%) were purchased from Sigma-Aldrich and used as received. The water used in the experiments was obtained from a Millipore water purification system with a resistivity of 18.2 MΩ·cm. The *m*-terphenyl isocyanide ligand, CNAr$^{Mes2}$, (where Ar$^{Mes2}$ = 2,6-(2,4,6-Me$_3$C$_6$H$_2$)$_2$C$_6$H$_3$) were prepared as previously described.[22]

**Surface Modification of AgNCs.** To perform the ligand exchange (LE) process, CNAr$^{Mes2}$ or CNXylyl ligands was added to 1 mL of AgNC dispersion to achieve a final concentration of 50 μM. The mixture was reactivated for 4 hours to complete the ligand exchange. After the reaction, the dispersion was washed twice to remove excess ligands. The resulting sediment was then re-dispersed in chloroform for use in further experiments.

**DFT Calculations.** To calculate the binding energies of CNAr$^{Mes2}$ and CNXylyl to Ag surfaces, Density Functional Theory (DFT) calculations were performed using the Quantum Espresso (QE) package,[59] employing the PBE exchange-correlation functional[60] with self-consistent dispersion correction[61] and the ultrasoft pseudopotentials of Dal Corso *et al.*.[62] The wavefunction kinetic energy cutoff was set to 60 Ry with a convergence tolerance of $1 \times 10^{-8}$ Ry, while the cutoff for the charge density was set to 480 Ry. A 15×15×15 K-point grid was used to efficiently sample the Brillouin zone. To capture the interactions of the isocyanide ligands with Ag, the (100) and (111) surfaces were constructed, consisting of a minimum of four layers of Ag atoms and a vacuum layer of at least 40 Å. Self-Consistent Field (SCF) calculations were performed on the isolated ligand, bulk Ag, and each of the slab geometries. The total energies of the DMP and Xylyl ligands on each of the Ag slabs were obtained, with the distance between the isocyanide carbon and the slab surface set to values ranging from 2 to 10 Å in 1 Å increments, followed by a finer scale of 0.1 Å increments around the energy minima. The binding energy was calculated using the following equation:

$$E_{binding} = E_{slab-ligand} - E_{slab} - E_{ligand}$$

**MD Simulations.** A 30 × 30 × 30 Å silver nanocube with exposed (100) faces, consisting of 2457 atoms, was placed in the center of a 100 × 100 × 100 Å$^3$ box. Ninety-eight CNAr$^{Mes2}$/CNXylyl ligands were added, and the entire system was solvated with 7,300 chloroform molecules using the Packmol package.[63] Classical molecular dynamics (MD) simulations were performed using the LAMMPS simulation engine.[64] The Ag atoms constituting the nanocube were described using the Embedded Atom/Finnis-Sinclair Method, with pairwise interactions developed by Ackland and coworkers.[19] The CNAr$^{Mes2}$ and CNXylyl ligands were primarily described with the GAFF force field,[65] while the C≡N bond stretch and ring torsions were treated using QMcorrected parameters from our previous work.[31] Chloroform molecules were described using parameters developed by Kamanth *et al.*.[66] The Ag–ligand van der Waals (vdW) interactions were obtained via Lorentz-Berthelot mixing rules,[67] assuming Ag vdW parameters from UFF.[68]

The van der Waals and real-space Coulomb cutoffs in the MD simulations were set to 10 Å. A cubic spline was applied to the vdW interactions to ensure smooth convergence and vanishing energies and forces at the cutoff (inner cutoff distance of 9 Å). Reciprocal space Coulomb interactions were computed using a particle-particle-particle-mesh solver with an error tolerance of 10$^{-6}$.[69] Each MD simulation was initiated with 500 conjugate gradient steps, followed by gradual heating to 5 K using 0.5 ns (500,000 steps with a 1 fs integration timestep) in the canonical ensemble (NVT; constant number of particles *N*, volume *V*, and temperature *T* = 298.15 K). A Nosé-Hoover thermostat was employed with a temperature relaxation window of 100 fs. Time integration was performed using the time-reversible, measure-preserving Verlet integrators derived by Tuckerman *et al.*.[70] After density equilibration, the system was simulated in the NVT ensemble for at least 5 ns of NVT dynamics.

**Simulated Raman Spectra.** To observe the C≡N stretching mode, simulated Raman spectroscopy was done using a modified form of the 2-Phase Thermodynamics code (2PT).[56] After 5 ns of NVT dynamics, three unique sampling groups of molecules were obtained by identifying isocyanide ligands that were either freely floating in solution, bound to a corner, or bound to a planar site. Next, the MD trajectory was generated from an additional 200 ps of NVT dynamics, with sampling occurring every 4 fs.

The vibrational power spectrum $\alpha_\omega^{vib}$ was obtained from the generated trajectory file:

$$\alpha_\omega^{vib} = \int_{-\infty}^{\infty} dt \, \langle V_{vib}(t) \cdot V_{vib}(0) \rangle \exp(i\omega t)$$

where *t* is time, *ω* is frequency, $V_{vib}$ is the vibrational component of the velocity of the species.

The IR spectra of neutral species ($\alpha_\omega^{IR}$) was generated from the fast Fourier transform (FFT) of the time correlation of molecular dipole moments ($D_{mol}$):

$$D_{mol} = \sum_1^n q \cdot \vec{r}$$

Where '*n*' is the number of atoms in a molecule, '*q*' and '$\vec{r}$' are charge and position vector of the atom, respectively, and

$$\alpha_\omega^{IR} = \frac{2\pi\omega(1 - \exp(-\beta\hbar\omega))}{3\hbar c V \mu_\omega} * \int_{-\infty}^{\infty} dt \, \langle D_{mol}(t) \cdot D_{mol}(0) \rangle \exp(i\omega t)$$

where *ℏ* is reduced Planck's constant, *c* is speed of light, *V* is the volume of the species under consideration, $\mu_\omega$ is the refractive index of the medium at frequency *ω* and *β* = 1/*kT*, *T* being the temperature and *k* the Boltzmann constant. $\mu_\omega$ is considered constant for simplicity.

The "unscaled" Raman spectra ($\alpha_\omega^{Raman}$) was generated by subtracting the IR spectrum from the vibration power spectrum of the species:

$$\alpha_\omega^{Raman} = \alpha_\omega^{vib} - 1 * \alpha_\omega^{IR}$$

**Synchrotron Infrared Nanospectroscopy (SINS).** SINS measurements were conducted at beamline 5.4 of the Advanced Light Source at Lawrence Berkeley National Laboratory. Infrared light was focused onto the apex of a Pt-coated AFM tip (NCH-Pt by Nanosensors) within a modified AFM system (Innova, Bruker).[37] Due to the nonlinear dependency of the near-field scattered signal on the tip–sample distance, tip oscillation induced higher harmonics (*nω*) in the scattered signal.[37] The second harmonic frequency (2*ω*) was detected using a lock-in amplifier to isolate the near-field signal from the far-field background.[37] A modified commercial FTIR spectrometer (Nicolet 6700, Thermo Scientific) was used to collect the infrared nanospectroscopy measurements. After obtaining AFM topography images of the sample surface, IR nanospectroscopy point measurements were performed at selected locations. The complex-valued near-field spectrum was derived from the Fourier transform of the interferogram, with the real (Re($v$)) and imaginary (Im($v$)) components represented as spectral amplitude $A(v)$ and phase $\phi(v)$, respectively. Near-field spectra were reported in the form of a normalized scattering phase, $\phi(v) = \phi_{sample}(v) - \phi_{reference}(v)$, using the bare Au-coated glass substrate as the reference. Spectral processing was performed using custom software developed at the beamline's end station. IR nanospectroscopy measurements were performed at multiple sites across the surfaces of several nanoparticles in each sample. At each location, spectra were collected repeatedly and averaged to improve the signal-to-noise ratio. Prior to acquiring each IR spectrum, the sample was allowed to reach thermal equilibrium, ensuring thermal drift was less than 5 nm/min. The IR signal was continuously monitored during acquisition to maintain precise tip positioning and measurement accuracy.

**Scanning Tunneling Microscopy (STM) and STM-Inelastic Electron Tunneling Spectroscopy (STM-IETS).** The STM and STM-IETS experiments were performed using a customized CreaTec lowtemperature STM operating at approximately 5 K and a base pressure of less than $1 \times 10^{-10}$ Torr. The Ag(111) substrate was cleaned by successive cycles of Ar$^+$ sputtering and thermal annealing. The electrochemically etched W tip was cleaned and sharpened by Ar$^+$ sputtering and thermal annealing, followed by conditioning through repeated poking on the Ag(111) surface until single-molecule resolution was achieved. The thermal stability of CNAr$^{Mes2}$ has been examined.[32] We dosed the CNAr$^{Mes2}$ ligands onto the clean Ag(111) surface at 5 K via thermal sublimation using a homemade Knudsen cell evaporator within the vacuum chamber. To promote surface diffusion of CNAr$^{Mes2}$ ligands, the sample was gradually warmed from 5 K to room temperature by removing it from the STM cryostat. Subsequently, the sample was cooled back down to 5 K for examination.

Topographic images were acquired in constant current mode by recording the $z$-position with feedback engaged and processed using Gwyddion.[71] The $d^2I/dV^2$ spectra were obtained by recording the second harmonic output of a lock-in amplifier while sweeping the bias voltage. A modulation of 3 mV (root mean square) at a frequency of 377 Hz was applied to the sample bias, with feedback turned off during bias sweeping.

## AUTHOR INFORMATION

**Corresponding Author**
*afrano@ucsd.edu

**Author Contributions**
[⊥]These authors contributed equally to this work
A.F., A.R.T., T.A.P., S.L., and J.S.F. conceived and led the project. The SINS experiments were performed by D.L., W.S., H.A.B., S.G.C., T.M., R.S., R.B., and Y.S.. Nanoparticle samples were prepared and characterized by Y.W., K.B., W.S., Y.S., and F.L.. DFT and MD simulations were performed by A.D., I.M., and G.R.. The STM experiments were performed by L.B. and S.L.. Data analysis was carried out by D.L., W.S., A.D., and L.B. The manuscript was written by A.F., D.L., A.R.T., T.A.P, and J.S.F. with input from all the coauthors.

**Notes**
The authors declare no competing financial interest.

## ACKNOWLEDGMENT
We thank the National Science Foundation, UCSD MRSEC DMR-2011924 for financial support. The authors acknowledge the use of facilities and instrumentation supported by the National Science Foundation through the UC San Diego Materials Research Science and Engineering Center (UCSD MRSEC) with Grant DMR-2011924.